# Swift Heavy Ion Tracks in Polymorphs of Titanium Dioxide: The Effects of Electron Beam and Crystal Structures

Xiaoting Jia[1,2,a)] Lijun Xu[2,3,a)], Ruslan A. Rymzhanov[4,5,*], Wenqiang Liu[2,3], Peipei Hu[2,3] Shengxia Zhang[2,3], Jiande Liu[2,3], Jian Zeng[2,3], Jie Liu[2,3], Pengfei Zhai[2,3,*]

[1]*College of Physics Science and Technology, Hebei University, Baoding 071002, China*

[2]*State Key Laboratory of Heavy Ion Science and Technology, Institute of Modern Physics, Chinese Academy of Sciences, Lanzhou 730000, China*

[3]*School of Nuclear Science and Technology, University of Chinese Academy of Sciences, Beijing 100049, China*

[4]*Joint Institute for Nuclear Research, Dubna 141980, Russia*

[5]*The Institute of Nuclear Physics, Almaty 050032, Kazakhstan*

[a)]X.J. and L.X. contributed equally to this work.

[*]Authors to whom correspondence should be addressed: rymzhanov@jinr.ru (Ruslan A. Rymzhanov), zhaipengfei@impcas.ac.cn (Pengfei Zhai)

**Abstract:**

Swift heavy ion (SHI) irradiation generates nanoscale damage in materials, offering opportunities for controlled modification of structural properties. Understanding the fundamental processes governing damage formation in altered regions remains a longstanding challenge. This work explores how crystal structure influences swift heavy ion-induced damage formation and evolution in rutile and anatase $TiO_2$. Combining transmission electron microscopy (TEM) and molecular dynamics simulations, we investigate how differences in atomic arrangement govern track morphology and structural recovery. Although irradiation produces amorphous tracks in both polymorphs, their morphologies differ considerably. In rutile, tracks develop pronounced diameter variations, forming hourglass-like structures, whereas in anatase they retain a relatively uniform cylindrical shape. Molecular dynamics simulations reveal that these differences arise from the distinct recrystallization behavior of the two crystal structures during rapid cooling, leading to different pathways of track evolution and recovery. Particular attention is paid to electron beam induced effects during TEM characterization, which can substantially modify the structure and apparent morphology of ion tracks. Prolonged electron beam exposure promotes the recrystallization of initially amorphous tracks and surface hillocks, accompanied by the development of Fresnel contrast. These findings highlight the importance of understanding beam-induced transformations for the reliable interpretation of TEM observations of radiation-damaged materials.

## 1. Introduction

When swift heavy ions (SHI) penetrate a target material, energy is primarily deposited through electronic energy loss, and once a threshold is exceeded, nanoscale damage regions, known as ion tracks, could be formed along the ion trajectory, typically with diameters of several nanometers.[1] The introduction of ion tracks with appropriate irradiation fluence can effectively modify the physical properties of materials or fabricate nano-functional materials. For example, SHI irradiation can significantly enhance the performance of YBCO superconducting materials, such as critical current density and irreversible magnetic field.[2,3] Sub-10 nm nanopores in $WO_3$ nanosheet can be directly produced with single SHI, and the size of the created nanopores could be well-controlled by changing electronic energy loss of SHI.[4] The characteristic timescale for ion track formation is generally recognized to be merely on the order of hundreds of picoseconds.[5,6] Accordingly, details of ion tracks, including their crystalline state, morphology, and dimensions, act as messengers that illuminate the spatiotemporal evolution of tracks under such extreme nonequilibrium conditions. These nanoscale ion tracks can be characterized using various techniques, including transmission electron microscopy (TEM), atomic force microscopy (AFM), Rutherford backscattering spectrometry in channeling mode (RBS/c), and small-angle X-ray scattering (SAXS).[1,7] Among these techniques, TEM is the only method capable of directly revealing both the morphology and phase state of ion tracks.

Ion tracks in the materials, such as $Y_3Fe_5O_{12}$,[8] $LiNbO_3$,[9] and zircon,[10] are amorphous and cylindrical, which suggest that the existing of transient thermal spike and subsequent quenching. Recently, the conical shape of ion tracks was found by using TEM in the near surface of rutile $TiO_2$,[11,12] yttria-stabilized zirconia (YSZ),[11] $Al_2O_3$,[11] and $CeO_2$.[13] The experimental study of ion tracks in rutile $TiO_2$ show that the morphology of ion tracks evolves with the track length or the sample thickness, and ion tracks exhibit Fresnel contrast as the focusing conditions of TEM changed.[12] High-resolution TEM (HRTEM) of ion tracks in rutile $TiO_2$ suggest that many 1-2 nm nanovoids formed inside the tracks are contributed to the Fresnel contrast.[11,12] The similar phenomena were also observed in $CeO_2$.[13] The combined effect of the ejection of molten phase and strong recrystallization ability are believed to be the main factors for these novel morphology formation of ion tracks.[12–15] Especially, the creation of through nanopores in $WO_3$ and $CaF_2$ nanosheets by single SHI irradiation provides a representative illustration that molten phase ejection and its liquid properties (viscosity and surface tension) play the key role.[4,16] Although the significance of recrystallization ability in ion track formation has been increasingly recognized,[12–14,17,18] the factors governing the efficiency of recrystallization is still not fully understood. Several potential factors governing the recrystallization ability of ion track in materials were proposed, e.g. the simplicity of lattice structure,[17] the diffusion velocity of oxygen atoms,[17] the ratio of cation radius,[19–21] density differences of crystalline and amorphous phases.[22] However, the existing studies mainly rely on molecular dynamics (MD) simulations, and the experimental investigations are relatively limited.

$TiO_2$ has three natural polymorphs, i.e. rutile, anatase and brookite. Among these, single crystal rutile and anatase $TiO_2$ are commercially available. Moreover, the thermodynamic parameters of these two polymorphs are close,[23] making them a promising material system for exploring the differences (especially the recrystallization ability) in track formation between polymorphs. On the other hand, it should be mentioned that the influence of electron beams during TEM characterization on ion tracks cannot be ignored. As for the ion tracks in $SrTiO_3$ (STO), two seemingly contradictory experimental results have recently been reported.[24–26] Ishikawa et al.[24] showed that the ion tracks in STO are crystalline with Fresnel contrast. In contrast, Gupta et al.[25,26] demonstrated that the initial ion tracks in STO are amorphous with uniform contrast, and these ion tracks readily change from amorphous to crystalline under electron irradiation. Evidently, as demonstrated by the above examples, the electron beam effect is likely to hinder our recognition and understanding of the intrinsic formation mechanism of ion tracks.

Therefore, in the present study two crystalline polymorphs (rutile and anatase) of $TiO_2$ were selected as research objects, and the interference induced by electron beams on ion tracks was carefully excluded. It was found that all of the ion tracks in both rutile and anatase $TiO_2$ are initially amorphous, but the morphology of ion track in rutile $TiO_2$ changes from cylindrical to dumbbell-like, then to hourglass-like morphologies with increasing the track length, while ion track in anatase $TiO_2$ show no length-dependent morphology. Combined with molecular dynamics simulations, it indicates that the distinct crystal structures of these two polymorphs are responsible for the different recrystallization abilities. These findings provide new insights into comprehensive understanding of ion track formation.

## 2. Results and Discussion

### 2.1. Electron-Beam Effects on Ion Tracks in Rutile and Anatase $TiO_2$

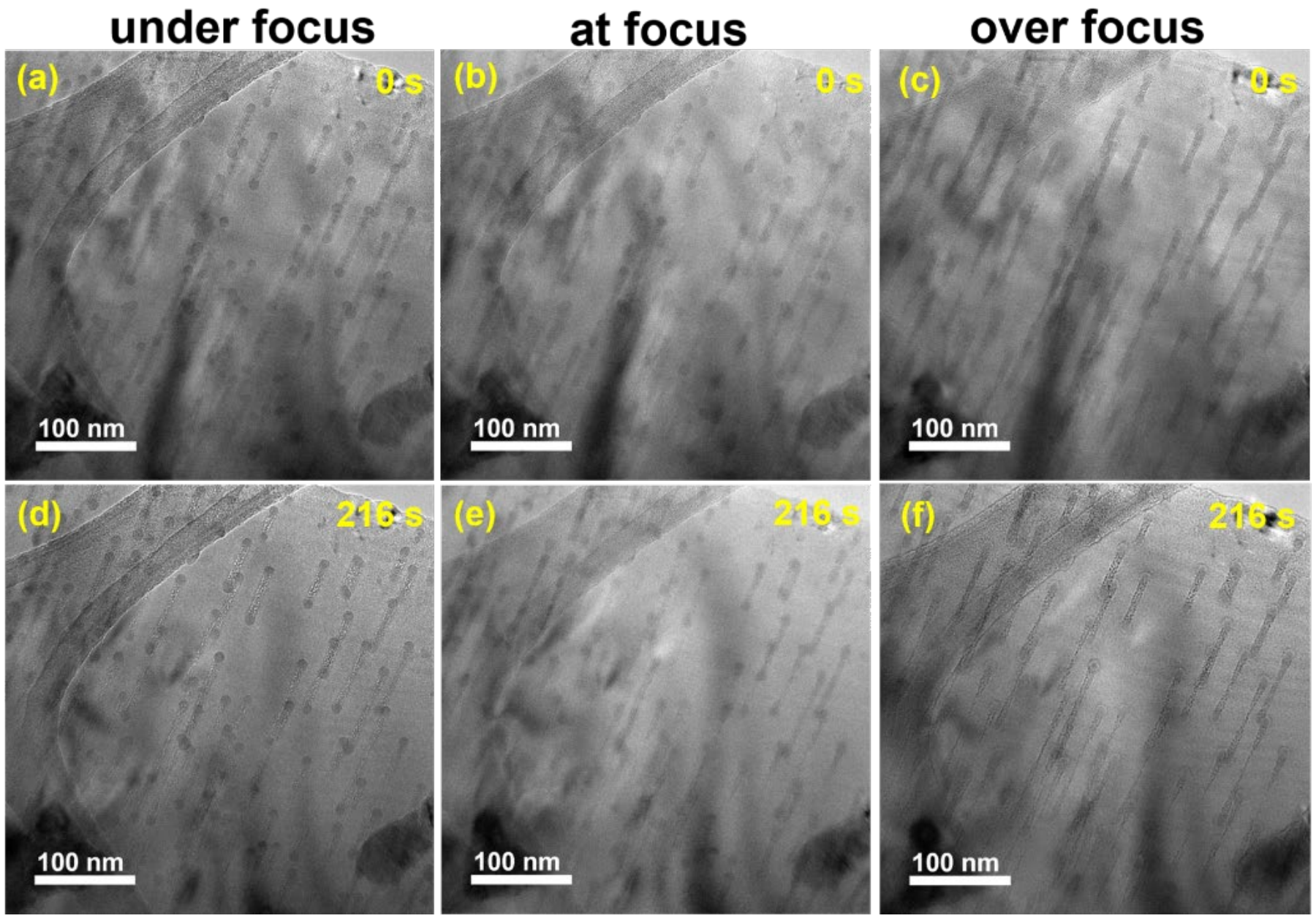

**Figure 1.** TEM images of ion tracks in rutile $TiO_2$ induced with 1128 MeV $^{181}Ta$ ions at the electronic energy loss of 34.7 keV $nm^{-1}$ and the incidence angle of 45°. Images (a)–(c) were acquired before continuous electron-beam irradiation, whereas images (d)–(f) were acquired after 216 s electron exposure with an electron dose rate of 400 e $Å^{-2}$ $s^{-1}$. Images (a)–(c) and (d)–(f) were recorded under underfocused, at-focused, and overfocused conditions, respectively.

Thanks to oblique-incidence irradiation, the whole ion tracks in rutile $TiO_2$ induced with 1128 MeV $^{181}$Ta ions could be observed in TEM (**Figure 1**a–c). A nano-hillock exhibiting circular contrast is visible at both ends of each ion track. It can be found that the geometric morphology of individual ion tracks in rutile $TiO_2$ varies with the track length: The short ion tracks exhibit a cylindrical morphology, while the long ion tracks appear hourglass-like shapes. In addition, the ion tracks with intermediate lengths show a dumbbell-shaped morphology. Importantly, no obvious Fresnel contrast was observed under either underfocused or overfocused imaging conditions. After the intrinsic ion tracks were recorded, the same region was continuously exposed to electron beam irradiation at a dose rate of 400 e Å$^{-2}$ s$^{-1}$ for 216 s to examine the potential influence of electron-beam effects on the ion tracks. As shown in **Figure 1**d–f, after prolonged electron-beam exposure Fresnel contrast appear inside these same ion tracks, which show the white core and dark fringe contrast at underfocused condition, whereas dark core and white fringe at overfocused condition. This inhomogeneous contrast indicates that the low-density nanoscale regions or nano-voids are formed. The same features of the ion tracks in rutile $TiO_2$ were reported in the previous work.[11,12] Unfortunately, they were thought to be the intrinsic fine structure of ion tracks in rutile $TiO_2$, [11,12] rather than the unintentional "artifact" caused by the electron beam. It should be noted that no obvious change occurs in the geometric morphology of the ion tracks in rutile $TiO_2$ upon prolonged electron beam exposure. Ion tracks in anatase $TiO_2$ are also sensitive to the prolonged electron beam exposure (**Figure S1**). To further verify the influence of the electron beam effects, top-view TEM characterization of ion tracks in rutile $TiO_2$ were performed (electron beam direction approximately parallel to the ion-track axis).

The influence of electron beam with different dose rates on ion tracks were investigated (**Figure 2**). In **Figure 2**a–c, the TEM images were obtained at a relatively low magnification of 310 kx (767 e Å$^{-2}$ s$^{-1}$). It can be found that the ion tracks in rutile $TiO_2$ are nearly circular and amorphous. The contrast of the tracks is uniform and no obvious Fresnel contrast is observed. At an intermediate magnification of 500 kx (2300 e Å$^{-2}$ s$^{-1}$), the ion tracks show the amorphous core around with crystalline halo (**Figure 2**d-f). The contrast is no longer uniform, but exhibit slight Fresnel contrast under different focus conditions. When the magnification was further increased to 820 kx (4620 e Å$^{-2}$ s$^{-1}$), within only ~1 s the initially amorphous ion tracks become almost crystalline, leaving obvious Fresnel contrast in the track core. These results indicate that, during conventional high-magnification TEM characterization (usually accompanied with high electron dose rate), even very short electron-beam exposure is able to modify the original structural state of ion tracks. This transform from amorphous state to crystalline one could be too fast to identify its intrinsic feature, if one tries to resolve the details of the ion tracks by simply increasing the magnification of TEM. On the basis of the above analysis, we suggest that the previously reported TEM observations of ion tracks in materials should be re-examined with consideration of potential electron-beam-induced effects. To demonstrate the process of recrystallization, continuous in-situ TEM imaging of the identical area was carried out under underfocused conditions with an intermediate electron dose rate.

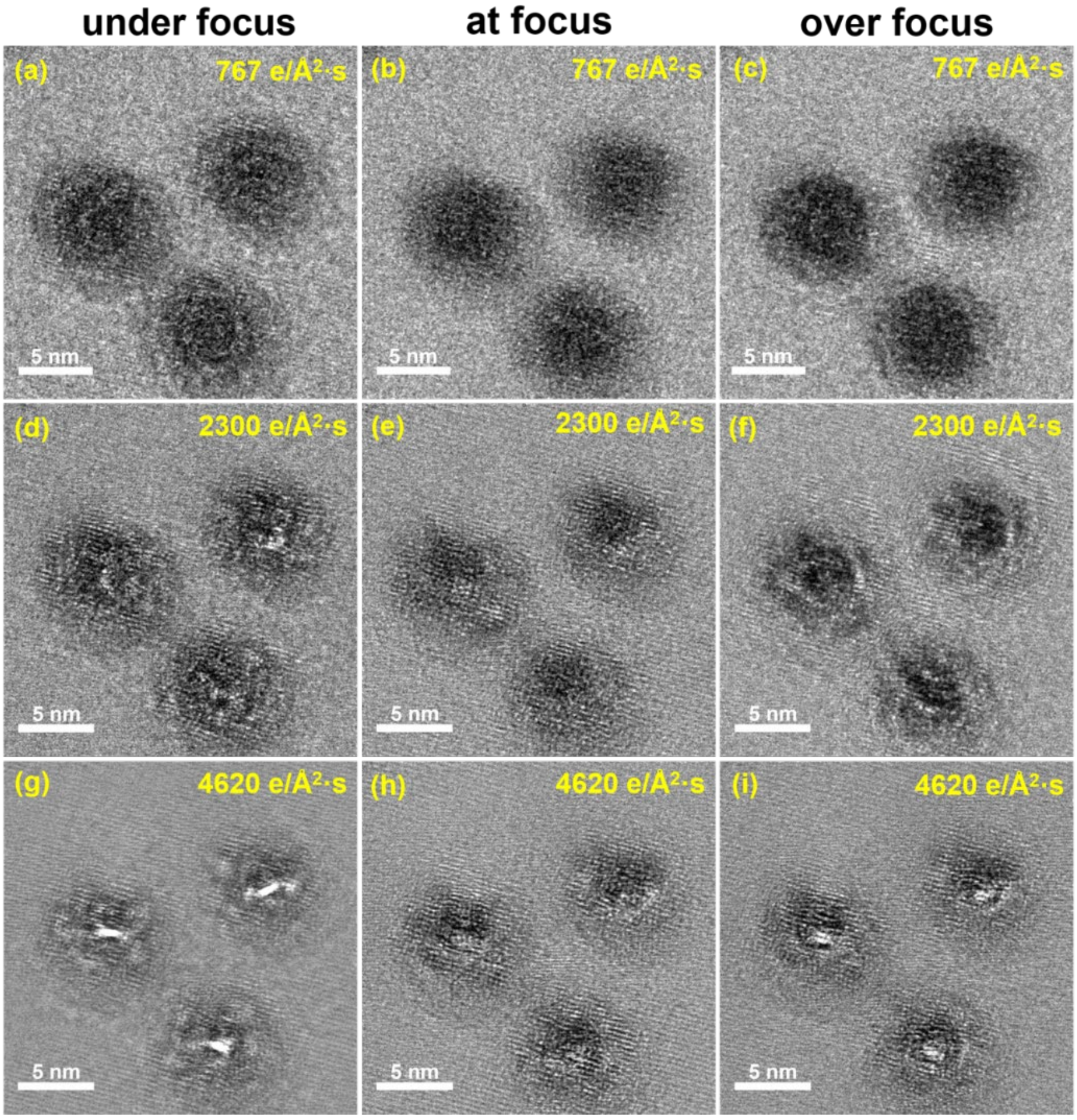

**Figure 2.** Top-view TEM images of ion tracks in rutile $TiO_2$ acquired with different electron dose rates and focus conditions. The ion tracks were produced by 2451 MeV $^{181}$Ta ions under normal incidence (30.0 keV nm$^{-1}$). (a–c) At the magnification of 310 kx (767 e Å$^{-2}$ s$^{-1}$), the tracks are nearly amorphous and the contrast is uniform. (d–f) At 500 kx (2300 e Å$^{-2}$ s$^{-1}$), the tracks remain amorphous cores with crystalline halos, and the Fresnel contrast start to appear. (g–i) At 820 kx (4620 e Å$^{-2}$ s$^{-1}$), the originally amorphous ion tracks almost completely recrystallize within 1s. Furthermore, the Fresnel contrast becomes more pronounced.

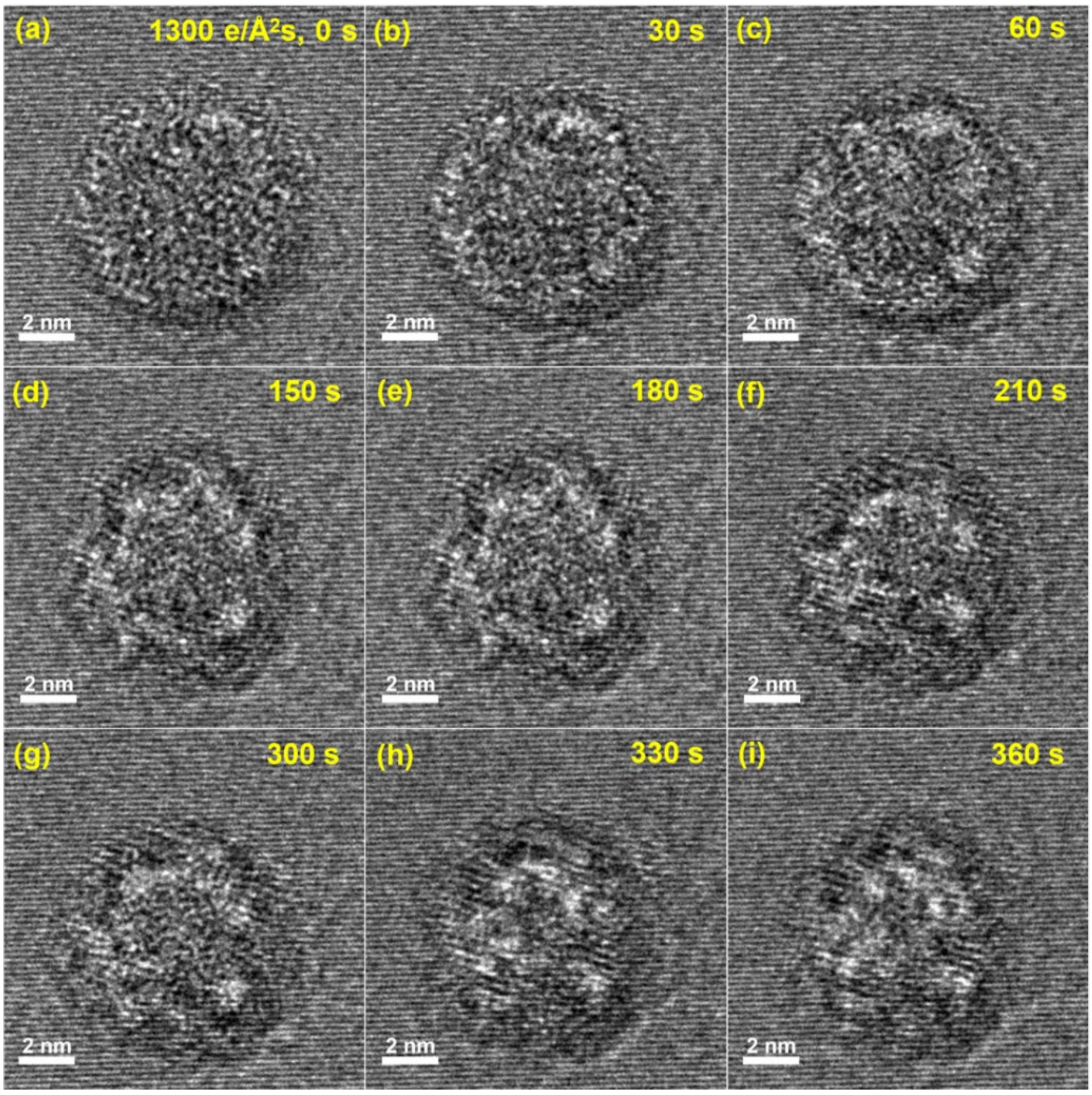

**Figure 3.** Time evolution of ion track in rutile $TiO_2$ during continuous electron-beam exposure under underfocused conditions. The ion track was produced by 2451 MeV $^{181}$Ta ion under normal incidence (30.0 keV nm$^{-1}$). The TEM images were acquired with an electron dose rate of 1300 e Å$^{-2}$ s$^{-1}$ before (a) and after electron exposure for 30 s (b), 60 s (c), 150 s (d), 180 s (e), 210 s (f), 300 s (g), 330 s (h), and 360 s (i).

**Figure 3**a shows that at the beginning the individual ion track is amorphous and circular with no Fresnel contrast. Following the continuous electron-beam exposure at a fixed electron dose rate (1300 e Å$^{-2}$ s$^{-1}$), weak white contrast gradually appears around the edge of the track. Between 150 and 300 s, this edge contrast intensified, and the boundary of the ion track become less distinct and also irregular. During 330–360 s of exposure, the Fresnel contrast shifts progressively toward the track core. It indicates that the recrystallization of amorphous ion track in rutile $TiO_2$ induced by electron beam start from the track edges, but the newly formed lattice fringes seem not fully consistent with those of the matrix. The ion

track in anatase $TiO_2$ under continuous electron-beam exposure (1300 e Å$^{-2}$ s$^{-1}$) shows the similar process which also recrystallizes starting from the boundary of the ion track (**Figure S2**). After prolonged electron beam exposure, the originally circular boundary is no longer distinct.

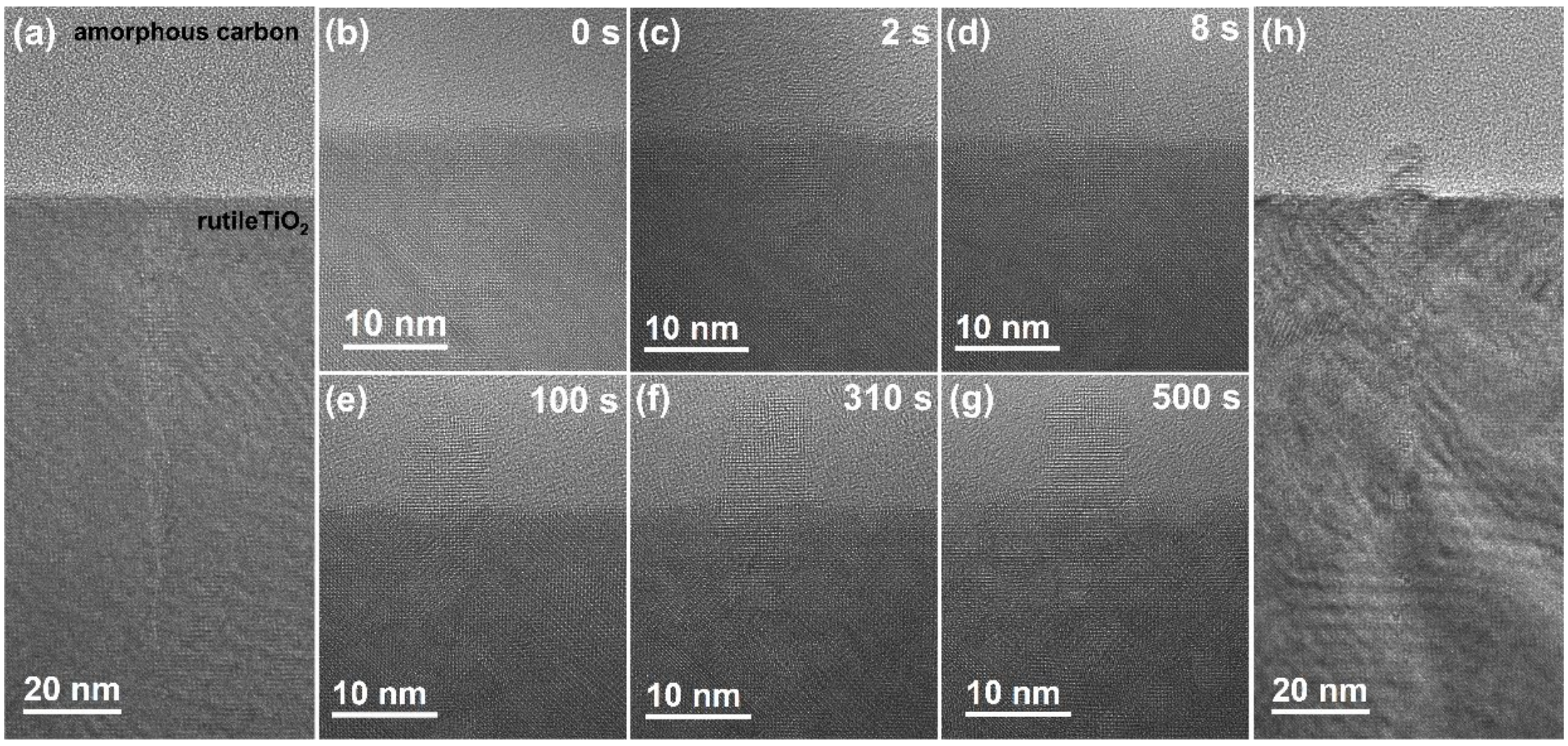

**Figure 4.** Cross-sectional TEM images of ion track near surface of rutile $TiO_2$. The ion track was produced by 2871 MeV $^{238}U$ ion under normal incidence (45.2 keV nm$^{-1}$). The TEM images were acquired with an electron dose rate of 4200 e Å$^{-2}$ s$^{-1}$ before (a) and after 500 s electron exposure (h), as well as the time evolution of electron-beam exposure for 0 s (b), 2 s (c), 8 s (d), 100 s (e), 310 s (f), 500 s (g). The amorphous carbon was deposited by magnetron sputtering to protect the hillock of ion track before preparing TEM sample using the focused ion beam.

Moreover, it is found that the ion-track hillock protruding from the rutile $TiO_2$ surface is also influenced by continuous electron-beam exposure (**Figure 4**). The morphology of the ion track near surface of bulk rutile $TiO_2$ induced with 2871 MeV $^{238}U$ ion (45.2 keV nm$^{-1}$) is conical with the cone length ~ 90 nm (**Figure 4**a). It is consistent with the previous study,[12] which suggest that the cone length of the ion track is positively correlated with electronic energy loss. At the beginning of TEM characterization, the hillock is amorphous, and nearly masked by the amorphous carbon layer (**Figure 4**a and **4**b). However, after exposure by electron beam with the electron dose rate of 4200 e Å$^{-2}$ s$^{-1}$ for only 2 s, the hillock starts to recrystallize from the surface (**Figure 4**c), indicating epitaxial crystallization templated by the surface lattice. The recrystallization rate is fast at the early stage, and slows down after ~100 s and above exposure of electron beam (**Figure 4**d-g). The recrystallization of hillocks stops after ~310 s. The final state of ion track near surface of bulk rutile $TiO_2$ is shown in **Figure 4**h. The whole region of rutile $TiO_2$ exposed to high dose of electron beam is influenced. Along the conical track, some faceted nanovoids could be found. Similar features were reported in the previous study.[12] The height of the crystalline hillock is about 10 nm. In previous work reported by O'Connell et al. [11], the hillock of ion track in rutile $TiO_2$ was believed to be originally crystalline. According to the present work, the electron-beam-induced effect on hillocks could have been ignored.

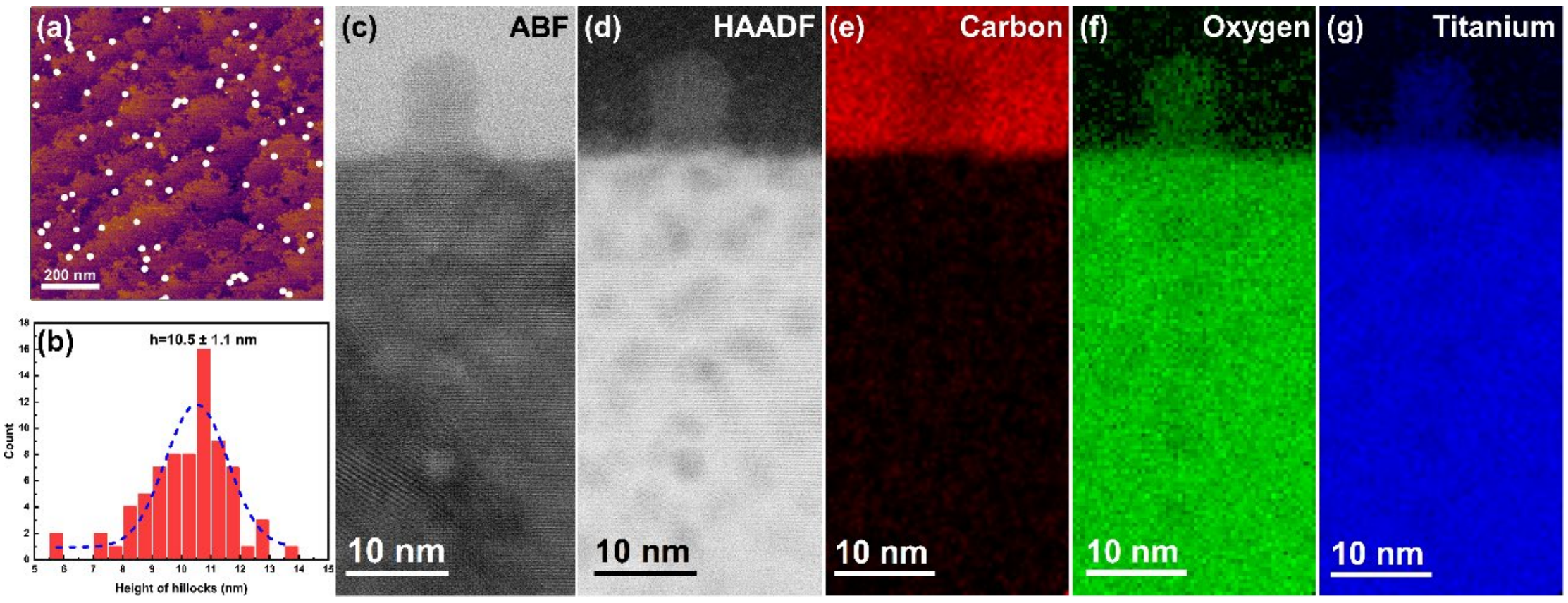

**Figure 5.** Ion track near surface of rutile $TiO_2$ characterized by AFM and STEM. The ion track was produced by 2871 MeV $^{238}U$ ion under normal incidence (45.2 keV $nm^{-1}$). (a) The AFM image was acquired after the bulk rutile $TiO_2$ irradiated with 2871 MeV $^{238}U$ ions without any additional processing. (b) The statistical analysis of the heights of hillocks was obtained from AFM measurement. The mean height and standard deviation were deduced by Gaussian fitting. The ABF-STEM image (c), HAADF-STEM image (d) and EELS elements mapping (e-g) of the ion track were shown to certify the elemental composition of the hillock.

**Figure 5** gives more information about the hillocks of ion tracks near surface of rutile $TiO_2$ irradiated with 2871 MeV $^{238}U$ ions (45.2 keV $nm^{-1}$). AFM measurement was performed directly after SHI irradiation without deposition of amorphous carbon. The randomly distributed hillocks are observed on the rutile $TiO_2$ surface using AFM (**Figure 5**a). Considering the tip convolution effect, we only performed statistics on the heights of hillocks. The mean height of the hillocks is measured to be 10.5 ± 1.1 nm (**Figure 5**b), which agrees well with the TEM results (**Figure 4** and **Figure 5**c). The same ion track as the **Figure 4**h was analyzed using annular bright field scanning TEM (ABF-STEM), high-angle annular dark-field scanning TEM (HAADF-STEM) and electron energy loss spectroscopy (EELS) mapping (**Figure 5**c-g). These results confirm that the hillock is $TiO_2$.

Based on these experimental observations, we analyze the possible mechanism of electron-beam-induced recrystallization. Because local recrystallization may be promoted by beam heating, the temperature rise induced by the 200 kV electron beam is estimated. The maximum temperature rise near the irradiated region can be expressed as follows:[27,28]

$$\Delta T = \frac{I}{\pi \kappa e}\left(\frac{\Delta E}{d}\right)\ln\frac{b}{r_0} \qquad (1)$$

where $I$ is the beam current, $\kappa$ is the thermal conductivity of the specimen, $e$ is the elementary charge, $b$ is the effective heat-sink radius, $r_0$ is the beam radius, and $\Delta E/d$ is the energy loss of electron per unit thickness. Because the electron energy loss is normally only a few eV $nm^{-1}$ and is much smaller than the incident electron energy, $\Delta E/d$ can be assumed to be constant. The total mass stopping power of 200 keV electrons in $TiO_2$ is 2.141 MeV $cm^2\ g^{-1}$, as obtained from the NIST ESTAR database.[29] Using the density of rutile $TiO_2$ (4.25 g $cm^{-3}$), the corresponding electron energy loss is estimated to be 0.91 eV $nm^{-1}$. Based on the beam current of 0.938 nA and the electron dose rate of 1300 e $Å^{-2}\ s^{-1}$ used in this work, the corresponding beam radius is estimated to be approximately 0.12 μm. Taking $\kappa$=0.084 J $K^{-1}\ cm^{-1}\ s^{-1}$,[23] $b$=1.5 mm and $r_0$=0.12 μm, the calculated temperature rise is approximately 0.31 K. Therefore, electron-beam heating is unlikely to be the underlying mechanism governing the observed recrystallization.

When electrons collide with lattice atoms, the maximum recoil energy can be expressed as follows:[30,31]

$$E = E_{\max}\sin^2(\theta/2) \qquad (2)$$

$$E_{\max} = E_0(1.02 + E_0/10^6)/(465.7A) \qquad (3)$$

where $E_0$ is the energy of the incident electron, $A$ is the atomic mass number of the target atom, and $\theta$ is the scattering angle between the incident and scattered electron trajectories. Under an accelerating voltage of 200 kV, the maximum energies that incident electrons can transfer to Ti and O atoms are 10.9 eV and 32.7 eV, respectively. According to the molecular dynamics simulations reported by Robinson et al.[32] the displacement threshold energies of Ti and O atoms in rutile $TiO_2$ are approximately 73 eV and 18 eV, respectively. These values indicate that the maximum transferred energy of TEM electrons to Ti atoms is much lower than the displacement threshold energy of Ti atoms. Thus, Ti atoms in the irradiated region are unlikely to be displaced from their lattice sites by elastic collisions with electrons and then migrate to defect sites to repair the damaged structure. In contrast, the transferred energy to O atoms is higher than the reported displacement threshold energy of O atoms, implying that oxygen related displacement cannot be completely excluded. Nevertheless, the transferred energy can exceed the displacement threshold only when electrons undergo high-angle elastic scattering with a scattering angle larger than approximately 95.7°, thereby possibly inducing knock-on displacement of O atoms. However, as shown in **Figure S3**, the effective knock-on displacement cross-section corresponding to this angular range is only about $6.90\times10^{-7}$ $a_0^2$ ($a_0^2$=$2.8002852\times10^{-21}$ $m^2$), accounting for approximately $3.40\times10^{-3}$% of the total elastic scattering cross-section. Consequently, although elastic displacement of O atoms is possible during 200 keV electrons irradiation, its actual occurrence probability is extremely low, making it unlikely to be the dominant mechanism for electron-beam-induced recrystallization of ion tracks.

Electron-beam-induced recrystallization of ion tracks is more likely related to inelastic ionization and electronic excitation. During electron-beam irradiation, electronic excitation may weaken the distorted Ti–O bonding network near the crystalline–amorphous interface between the track core and the surrounding crystal. This weakening can promote local bond rearrangement and defect recovery, possibly involving atomic hopping or rotation of $TiO_6$ polyhedra. At the initial stage, recrystallization may preferentially occur at the crystalline–amorphous interface, where the surrounding crystal can serve as a structural template for epitaxial regrowth. With continuous electron-beam irradiation, the accumulated energy deposition from electron excitation and ionization may become sufficient to promote recrystallization even within the fully amorphous track core, where no pre-existing crystalline seeds are available for growth.[33] Since electron excitation involves electrons transitioning to higher atomic energy levels and depends on the electron dose,[34] the TEM equipped with a K3 direct-electron detector enables efficient signal acquisition under low-dose conditions, allowing one to obtain the intrinsic ion tracks minimizing radiation damage of electrons.

## 2.2. Comparison of Ion Tracks in Rutile and Anatase $TiO_2$

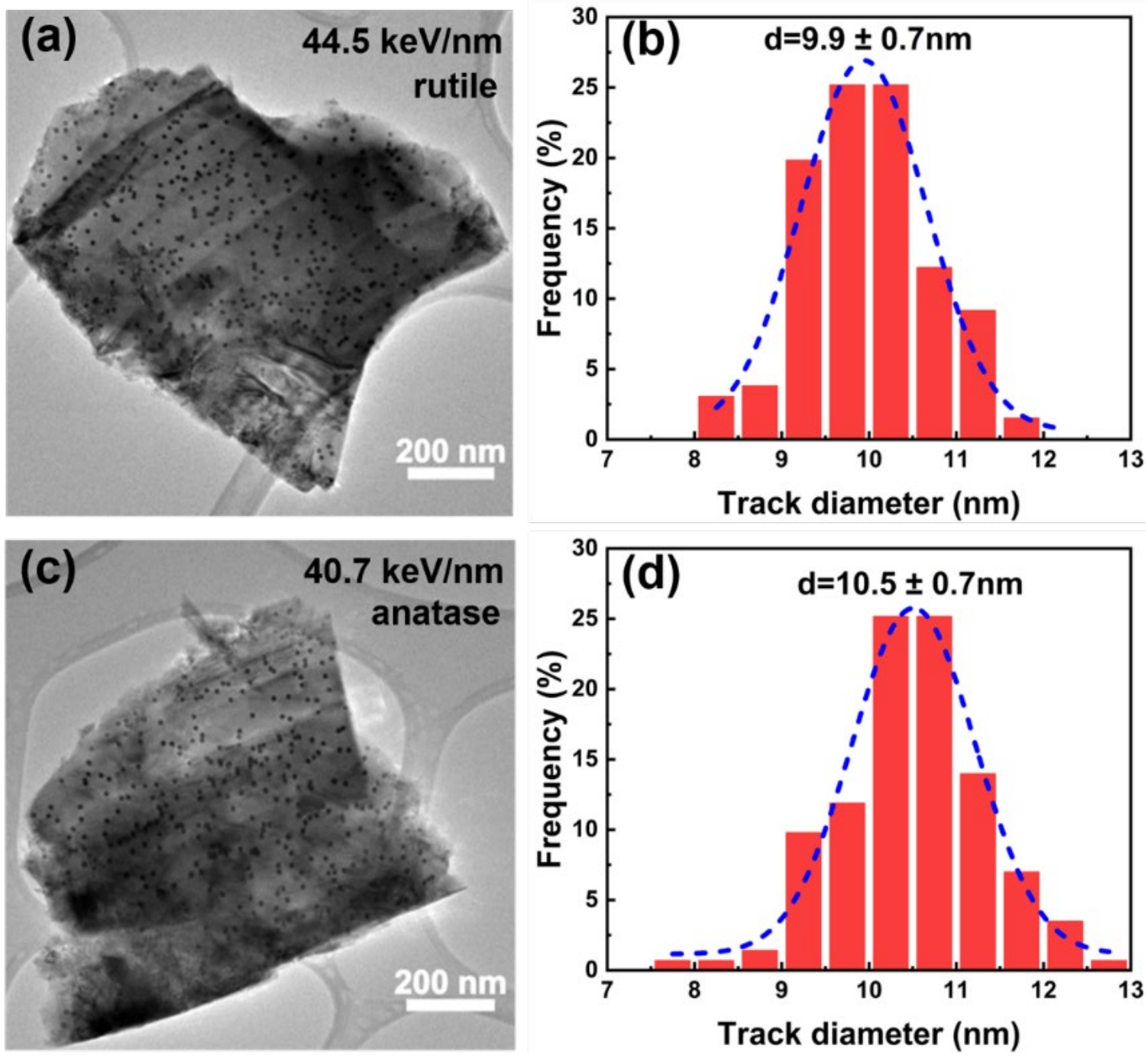

**Figure 6.** TEM images of $TiO_2$ in the rutile (a) and anatase $TiO_2$ (c) irradiated with 3040 MeV $^{238}U$ ions under normal incidence. The corresponding electronic energy losses are 44.5 keV $nm^{-1}$ for rutile and 40.7 keV $nm^{-1}$ for anatase, respectively. (b) and (d) show the Gaussian fitting of ion-tracks diameters measured from the TEM images.

Having clarified the influence of electron-beam effects on rutile ion tracks, we next compare the ion tracks in rutile and anatase $TiO_2$ under conditions of minimal beam damage. **Figure 6** shows the low-magnification TEM images of ion tracks in rutile (**Figure 6**a) and anatase $TiO_2$ (**Figure 6**c) produced by 3040 MeV $^{238}U$ ions under normal-incidence irradiation, with the electronic energy losses of 44.5 keV $nm^{-1}$ and 40.7 keV $nm^{-1}$, respectively. The ion tracks in both polymorphs of $TiO_2$ exhibit nearly circular profiles and are uniformly distributed. The measured areal density of ion tracks is generally consistent with the applied ion fluence. To quantitatively compare the track sizes in the two polymorphs, the diameters of more than 100 individual ion tracks were measured. The results show that the track diameters in rutile and anatase $TiO_2$ are close, with values of 9.9±0.7 nm and 10.5±0.7 nm, respectively (**Figure 6**b and d). These low-magnification TEM results indicate that the ion track diameters in the two phases are comparable within experimental uncertainty. To further reveal the phase-state changes of the ion track, HRTEM images and the corresponding fast Fourier transform (FFT) results are analyzed below.

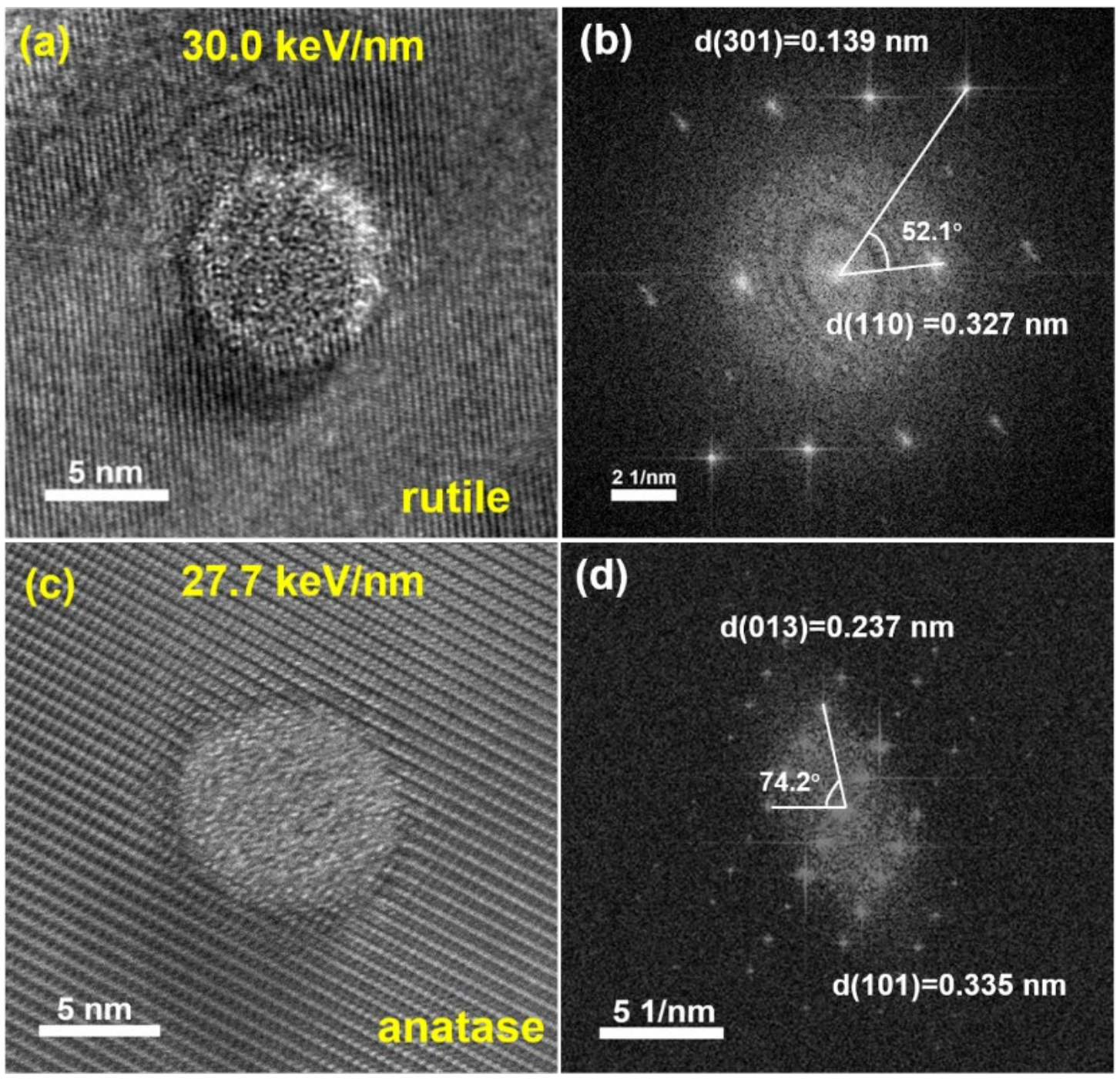

**Figure 7.** HRTEM images of ion tracks in rutile (a) and anatase $TiO_2$ (c) produced by 2451 MeV $^{181}$Ta ions and 2375 MeV $^{181}$Ta ions under normal incidence, respectively. The corresponding electronic energy losses are 30.0 keV $nm^{-1}$ for rutile and 27.7 keV $nm^{-1}$ for anatase. (b) and (d) show the FFT patterns obtained from the full fields of view in (a) and (c), respectively. The indexed diffraction spots in the FFT patterns indicate that the surrounding crystalline matrices in both rutile and anatase $TiO_2$ remain intact, confirming that the TEM observations reflect the intrinsic lattice structure outside the ion tracks.

**Figure 7** shows the HRTEM images of ion tracks produced by 2451 MeV $^{181}$Ta ions in rutile $TiO_2$ and 2375 MeV $^{181}$Ta ions in anatase $TiO_2$, together with the corresponding fast Fourier transform (FFT) patterns. Clear and well-defined diffraction spots can be observed in the FFT patterns, indicating that the crystalline matrix surrounding the tracks remains well ordered. In contrast, no discernible lattice fringes are observed within the track regions, suggesting the loss of long-range crystalline order and the formation of amorphous features. These observations indicate that ion tracks in both $TiO_2$ polymorphs consist of localized amorphous damage regions embedded within an intact crystalline matrix. Based on these results, the full-length geometric morphologies of ion tracks in two polymorphs under oblique-incidence irradiation is further compared below.

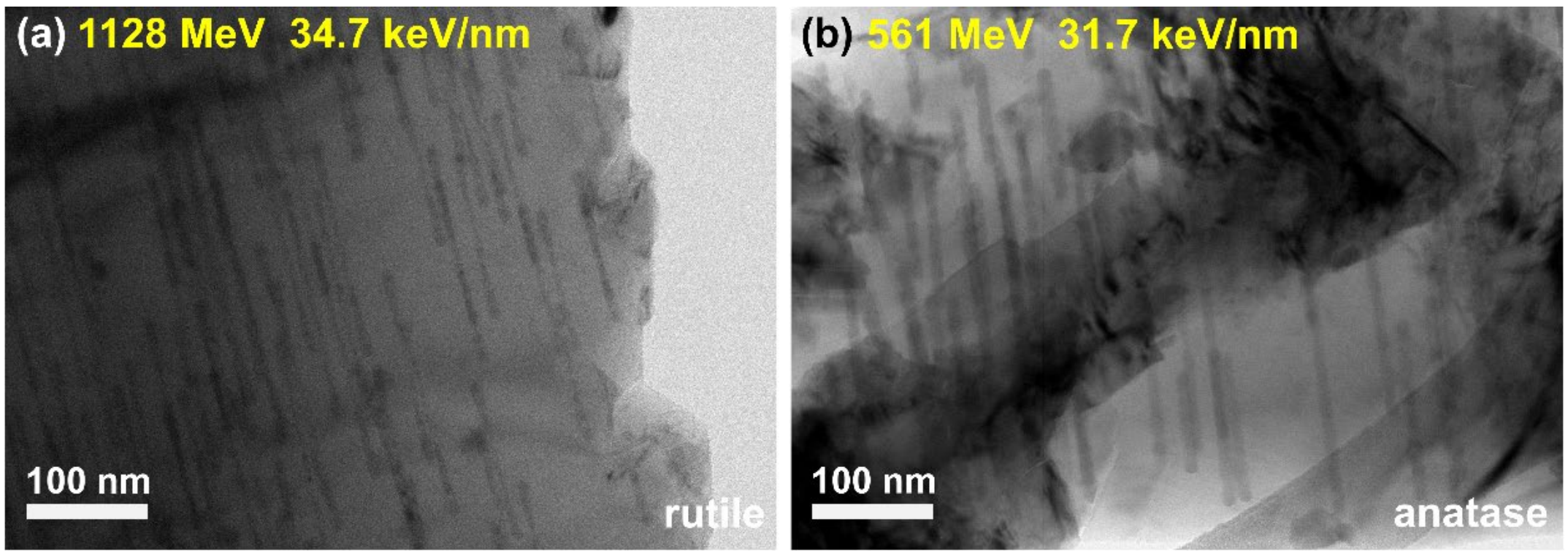

**Figure 8.** TEM images of ion tracks in rutile (a) and anatase $TiO_2$ (b) irradiated with 1128 MeV $^{181}Ta$ (34.7 keV $nm^{-1}$) and 561 MeV $^{181}Ta$ (31.7 keV $nm^{-1}$) under an incidence angle of 45°.

**Figure 8** shows the TEM images of elongated ion tracks in rutile and anatase $TiO_2$, which were irradiated with 1128 and 561 MeV $^{181}Ta$ ions under an incidence angle of 45°. The corresponding electronic energy loss are 34.7 and 31.7 keV $nm^{-1}$, respectively. As a comparison, the geometric morphology of the ion tracks in rutile $TiO_2$ evolves significantly with the track length (**Figure 1** and **Figure 8**a), whereas the ion tracks in anatase $TiO_2$ remain cylindrical morphology regardless of the track length. It indicates that the crystal structure is probably responsible for this distinct discrepancy in geometric morphology of ion tracks.

## 2.3 Molecular Dynamics Simulations of Ion Track Formation in Rutile and Anatase $TiO_2$

To understand the morphological differences observed experimentally between the two polymorphs, MD simulations were performed to investigate the structural response of rutile and anatase $TiO_2$ to ion irradiation. The damage induced by 2375 MeV Ta ions was studied in thin films with thicknesses of 60 and 120 nm. **Figures 9**a and **9**b show the formation of cylindrical amorphous ion tracks in rutile and anatase, with diameters of approximately 8 and 8.8 nm, respectively. The simulated track dimensions and morphologies are in reasonable agreement with the experimental results presented in **Figure 7**.

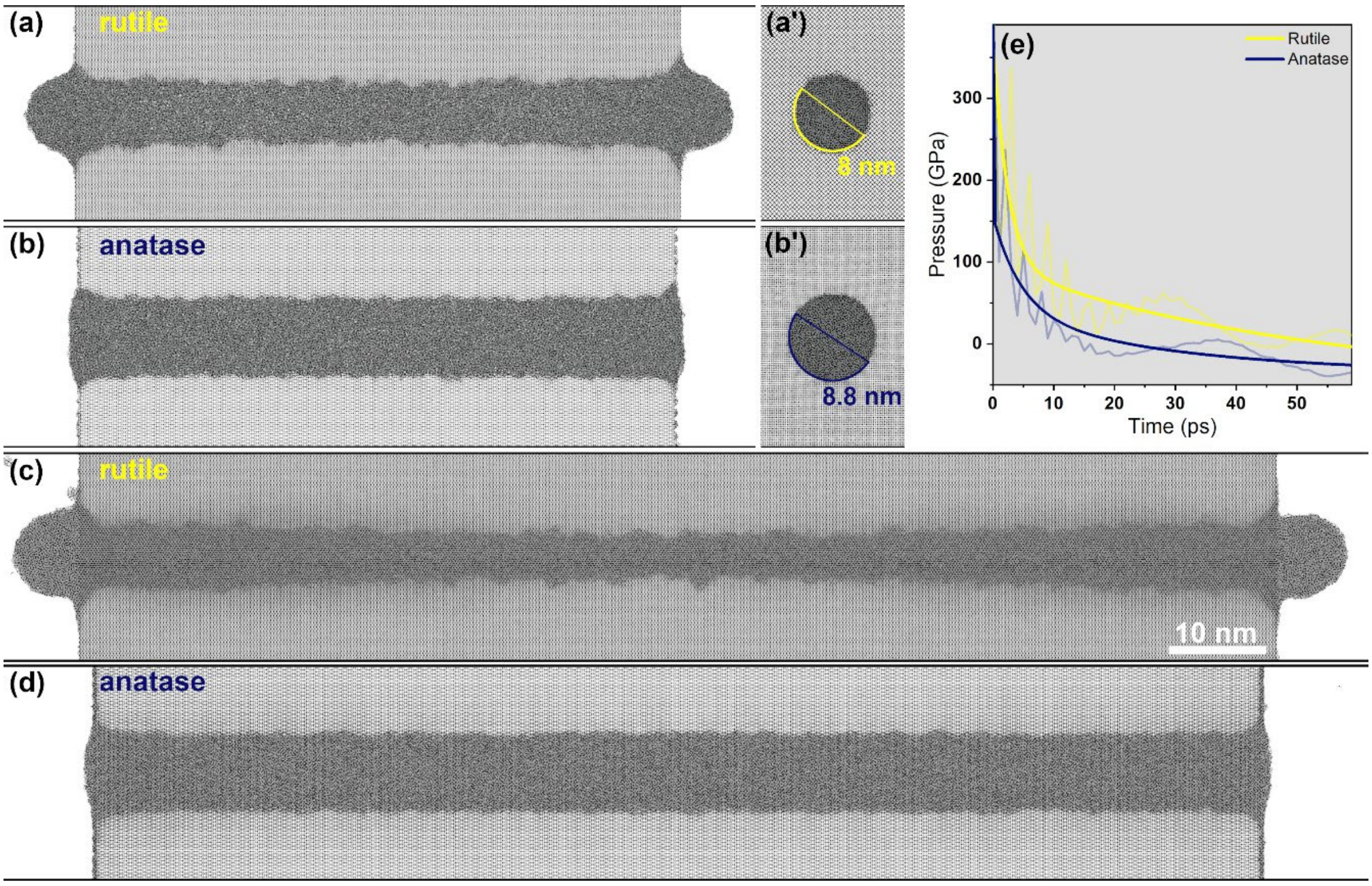

**Figure 9.** Molecular dynamics simulations of ion tracks formed by 2375 MeV Ta ions in rutile and anatase $TiO_2$. (a, b) Cross-sectional views of the damaged regions in 60 nm rutile and anatase films, respectively. (a′, b′) Views of the corresponding tracks images in planar geometry, with the diameters indicated. (c, d) Cross-sectional views of the tracks formed in 120 nm rutile and anatase films. (e) Time evolution of the pressure in the central region of the films during ion-track formation.

The MD simulations also reveal the formation of pronounced surface protrusions (hillocks) in rutile, whereas no comparable hillocks are observed in anatase. The pressure evolution in the central region of the films (**Figure 9**e) provides insight into the origin of this contrasting behavior. Rutile exhibits higher pressures throughout the simulated time interval, whereas anatase shows substantially lower pressures and faster relaxation toward negative values. At approximately 15–20 ps, the pressure difference between the two polymorphs reaches nearly an order of magnitude. This behavior can be attributed to the differences between the density of amorphous $TiO_2$ (3.87 g/cm$^3$) and those of crystalline rutile (4.33 g/cm$^3$) and anatase (3.89 g/cm$^3$), as calculated using MD simulations. Assuming the same properties of the protruding $TiO_2$ melt, such as viscosity and surface tension, the larger density difference between amorphous and crystalline rutile results in greater pressure accumulation, which may contribute to hillock formation. In contrast, the smaller density difference in anatase is consistent with more efficient stress relaxation and the absence of pronounced surface protrusions.

The simulated hillock morphology is in qualitative agreement with the experimental observations (**Figure 4**). Moreover, the simulated surface nanostructures appear to be amorphous. The ion tracks formed in the 120 nm films exhibit distinct morphologies in rutile and anatase. Anatase shows the formation of a continuous amorphous region following ion impact, similar to that observed in the 60 nm film. In contrast, the tracks in rutile exhibit an hourglass-like morphology, with a wider region near the surface and a narrower region in the middle of the film. This morphology is consistent with the TEM observations presented in **Figure 8**. In a previous study,[15] it was shown that the presence of a free surface and material ejection during track formation in $CeO_2$ can result in the formation of conical tracks. **Figure S4** compares the structural relaxation kinetics of rutile and anatase $TiO_2$, revealing that ion impact initially produces molten regions of approximately the same diameter in both polymorphs. During subsequent cooling and relaxation, the disordered region in rutile shrinks; however, this process is nonuniform along the ion trajectory. It was proposed that the ejection of molten material onto the surface results in a local depletion of matter in the near-surface region,[15] which may suppress recrystallization. This effect is most pronounced near the surface and gradually decreases with depth, resulting in conical or hourglass-like track morphologies. On the other hand, relaxation of the ion induced excitation in anatase does not show any recovery in the vicinity of an ion trajectory, thus molten region quenches in an amorphous state forming continuous amorphous track.

The conically shaped tracks in rutile are, in our opinion, clear indirect indication of the presence of recrystallization processes in rutile and their absence in anatase. The comparison of the response of two targets of the same chemical composition but different crystalline structures give a unique chance to uncover the crystallization driving forces in various materials. Previous study of tracks formation in MgO, $Al_2O_3$ and $Y_3Al_5O_{12}$ connected the recrystallization efficiency with the complexity of the structure of a target as well as with the mobility of atoms in molten state.[17]

Classical crystallization theory assumes that the crystal growth is described as a thermally activated process involving atomic transitions across the solid-liquid interface. According to this model, named Wilson–Frenkel (WF) model,[35-37] the interface velocity is governed by the competition between atomic attachment to the crystalline phase and detachment into the liquid phase. Thus, the crystallization and melting rates are determined by the corresponding activation barriers in the free-energy landscape. Specifically, the crystallization rate is governed by the activation energy for atomic attachment to the crystalline phase, $\Delta E_a$, whereas the reverse process involves an additional contribution associated with the chemical potential difference between the liquid and solid phases, $\Delta\mu$. The competition between these two processes determines the net interface growth rate *R(T)* dependence on temperature *T*, which can be expressed as

$$R(T) = k(T)\left[1 - e^{-\frac{\Delta\mu(T)}{k_B T}}\right] \quad (4)$$

$$k(T) = k_0 e^{-\frac{\Delta E_a}{k_B T}} \quad (5)$$

where $k(T)$ is known as the kinetic factor, $k_0$ is the kinetic prefactor related to the frequency of atomic transitions and $k_B$ is the Boltzmann constant. In the classical formulation of the WF model, the activation energy for crystallization is assumed to be equal to the self-diffusion activation energy in the liquid $\Delta E_a = \Delta E_d$. This assumption is based on the premise that atoms joining the crystalline phase undergo the same diffusion processes as those governing atomic mobility in the bulk liquid. However, the WF model often fails to quantitatively reproduce crystallization rates obtained from molecular dynamics simulations and experiments. This discrepancy has commonly been attributed to a reduction in atomic mobility in the supercooled liquid near the solid–liquid interface, which would result in an effective activation energy exceeding the bulk liquid self-diffusion barrier $\Delta E_a > \Delta E_d$.

MD simulations of two-phase systems were used to determine the crystal growth rates of rutile and anatase as a function of interface temperature (**Figure 10**a). The crystallization rate of rutile was found to be nearly an order of magnitude higher than that of anatase, consistent with the differences in swift heavy ion track morphology observed by TEM. A zero-growth rate indicates that crystallization and melting occur at equal rates and corresponds to the melting temperature. With the interatomic potential used here, the melting temperature of the MD system initially containing anatase was approximately 2350 K. The crystal-growth curves were fitted using Eq. (4). Because the classical WF model uses the self-diffusion activation energy, $\Delta E_d$, this parameter was determined from an Arrhenius plot of the diffusion coefficient (**Figure 10**b). The diffusion coefficient of atoms in liquid $TiO_2$ was found to be independent of the initial crystalline phase from which the liquid system was generated. We also assumed that the chemical-potential difference between the liquid and solid phases in Eq. (4) is equal to the corresponding free-energy difference ($\Delta\mu = \Delta G$). The free-energy difference between the crystalline and liquid phases (**Figure 10**c) was calculated following the procedures described in References [38,39].

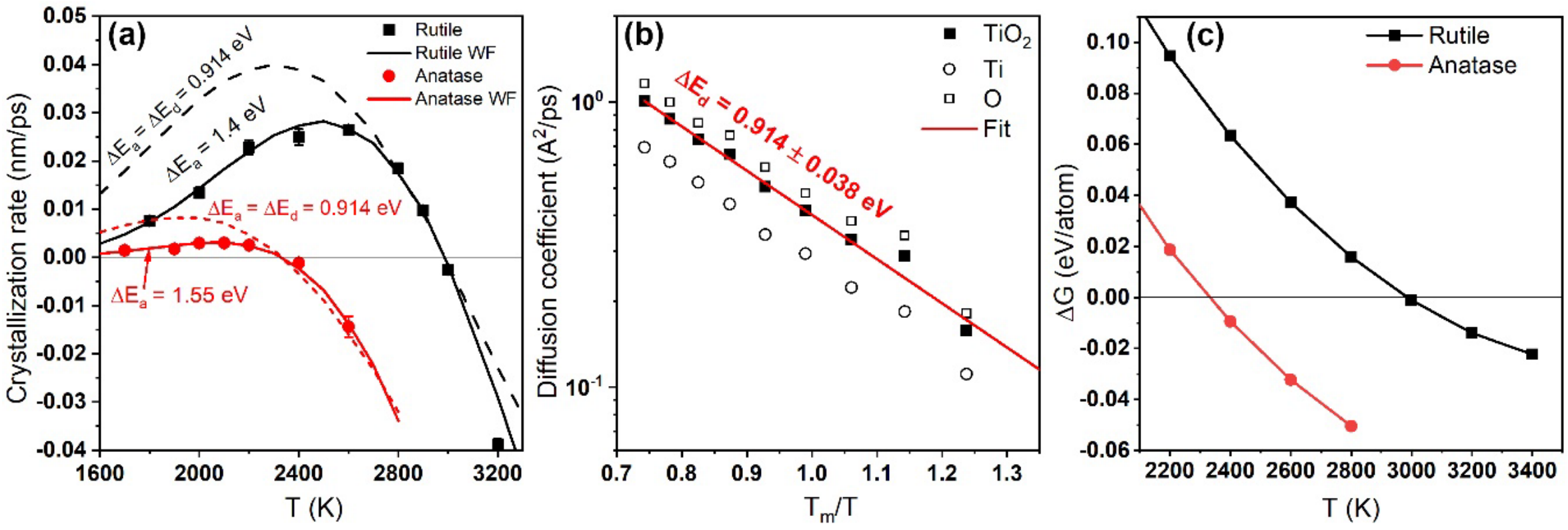

**Figure 10.** (a) Crystallization rate of rutile and anatase as a function of temperature, showing the Wilson–Frenkel model fits for the two polymorphs. (b) Arrhenius plot of the liquid-phase diffusion coefficient. The solid line represents fit to the data to obtain the diffusion activation energy $\Delta E_d$. (c) Free-energy difference $\Delta G$ between the crystalline and liquid phases as a function of temperature. The thermodynamic melting temperature is defined by $\Delta G=0$.

The dashed lines in **Figure 10**a show that using the liquid self-diffusion activation energy, $\Delta E_d$ = 0.914 eV, as the crystallization activation barrier, $\Delta E_a$, in Eq. (4) does not reproduce the temperature dependence of the growth rate. Fitting the MD data with Eq. (4) (solid lines in **Figure 10**a) yields crystallization activation energies of 1.40 eV for rutile and 1.55 eV for anatase. The fits assume the same kinetic prefactor, $k_0$ = 90.3 nm/ps, for both phases. The difference between $\Delta E_a$ and $\Delta E_d$ is commonly attributed to interface-induced ordering. [37] Such ordering has been observed in the liquid near the interface during crystal growth, when the solid-liquid interface is out of equilibrium. It appears to reduce liquid-atom mobility by increasing the crystallization activation barrier, $\Delta E_a$, and thereby slowing crystal growth. The extent of this ordering also depends on crystal orientation and interface roughness.

These results suggest that the differences in SHI track morphology between rutile and anatase may be related to their recrystallization kinetics. Rutile tends to recrystallize on the short timescales of track formation, whereas recovery in anatase is much slower and is therefore difficult to observe within the timeframe of SHI damage generation. The ability of the $TiO_2$ polymorphs to recover their initial structures may depend on the following factors:

(1) The difference in free energy between each crystalline phase and the liquid makes the liquid-to-rutile transformation more energetically favorable, resulting in a higher crystallization rate. Because the same interatomic potential is used for both polymorphs, this difference is likely governed primarily by potential energy minima landscape depending on crystal structure.

(2) The kinetic factors for crystallization can differ between rutile and anatase at the liquid-solid interface. If the self-diffusion activation energy $\Delta E_\mathrm{d}$ and kinetic prefactor $k_0$, are assumed to be the same for liquid $TiO_2$ in both cases, the observed difference in $k(T)$ may arise mainly from differences in interface-induced ordering in the two phases, which is most likely depends on crystal structure.

## 3. Conclusions

In this work, we re-examined the characterization of ion tracks in rutile $TiO_2$ by considering the influence of TEM electron-beam effect. The results show that the Fresnel contrast observed in rutile tracks is not an intrinsic feature of the as-formed tracks, but is induced during electron-beam exposure. After minimizing this beam-induced effect, the ion track morphologies in rutile and anatase $TiO_2$ were compared. Rutile tracks exhibit a clear length-dependent evolution along the ion trajectory, changing from cylindrical to dumbbell-like and finally to hourglass-like shapes, whereas anatase tracks remain predominantly cylindrical. This difference indicates that track morphology evolution is mainly governed by crystal structure variations.

## 4. Methods section

*Materials and Sample Preparation*: The experimental samples used in this study were (110)-oriented rutile single crystals and (001)-oriented anatase single crystals purchased from KJ-MTI Co., China. For TEM characterization, both types of single crystals were wet-ground in an agate mortar with ethanol before irradiation. After ultrasonic dispersion for 30 min, the supernatant was dropped onto lacey carbon film-coated copper/gold TEM grids to prepare TEM samples.

*Swift Heavy Ion Irradiation*: Swift heavy ion irradiation experiments were performed at the Heavy Ion Research Facility in Lanzhou (HIRFL). All irradiations were conducted at room temperature under high-vacuum conditions (~$5\times10^{-5}$ Pa). In the irradiation experiments, 13.5 MeV/u $^{181}$Ta, 16.0 MeV/u $^{181}$Ta and 15.5 MeV/u $^{238}$U ions were used. To investigate the effect of electronic energy loss on ion track formation, aluminum degraders with different thicknesses were used to change the incident ion energy and the corresponding electronic energy loss in the materials. Meanwhile, to compare the evolution of track morphology with ion penetration depth in rutile and anatase phases, 45° oblique irradiation experiments were also conducted. The applied ion fluence was set to approximately 1-$5\times10^{10}$ ions cm$^{-2}$ to avoid ion track overlapping. The ion flux was generally controlled below $10^8$ ions cm$^{-2}$ s$^{-1}$ to avoid macroscopic thermal effects during ion irradiation. The ion fluence and flux were monitored in real time using a secondary electron detector (18 μm Al in total) calibrated with a Faraday cup, and the uncertainty of the fluence measurement was usually ~10-20%. The electronic energy loss in the materials were calculated using the SRIM program. Detailed experimental parameters, including ion species, degrader thickness, incident angle, ion energy, material species, electronic energy loss and track diameter, are summarized in **Table S1**. For all of the irradiation experiments, the electronic energy loss is about two or three orders of magnitude higher than the nuclear energy loss. Thus, the effects of nuclear energy loss could be neglected.

*AFM Characterization*: Tapping mode AFM (Asylum Research, USA) was used to characterize the hillocks on the surface of the irradiated bulk rutile $TiO_2$ with 2871 MeV $^{238}$U ions at ambient atmosphere. The type of the AFM tip is Arrow-UHF (NanoWorld, Switzerland) with cantilever resonance frequencies of about 1.4 MHz. The height of the hillocks was analyzed using the built-in analysis software.

*TEM Characterization*: TEM samples characterization was mainly performed using a Talos F200X transmission electron microscope with the accelerating voltage of 200 kV. Bright-field transmission electron microscopy and high-resolution transmission electron microscopy (HRTEM) were employed to characterize the morphology and phase state of the ion tracks. To analyze the Fresnel contrast of the ion tracks, images of the same region were acquired under underfocused, at-focused, and overfocused conditions, respectively. To investigate electron-beam-induced structural evolution, the same region was continuously exposed to the electron beam, and sequential TEM images were acquired under exposure time and electron dose rate. To minimize electron-beam-induced modification of the intrinsic track structure, images were acquired immediately after the target magnification was reached. Fast Fourier transform was performed on the HRTEM images, followed by diffraction-spot indexing, to identify the crystalline phases of the matrices surrounding the ion tracks and to evaluate their degree of lattice ordering. Selected low-dose observations of the intrinsic ion track morphology were also performed using a Talos F200C transmission electron microscope (200 kV accelerating voltage) equipped with a K3 direct-electron detector. For the bulk rutile $TiO_2$ irradiated with 2871 MeV $^{238}U$ ions, after AFM characterization ~45 nm amorphous carbon layer was deposited on the sample using magnetron sputtering to protect the hillocks on the surface for TEM sample preparation by focused ion beam.

*Molecular dynamics simulations*: The electronic excitation induced by swift heavy ions was simulated using the TREKIS-3 Monte Carlo model,[40,41] which employs an event-by-event asymptotic trajectory algorithm to describe charged-particle interactions in solids. The model accounts for target ionization and the subsequent transport of primary and secondary electrons and holes, electronic cascades, core-hole radiative and Auger decays, photon absorption, valence-hole redistribution, and energy transfer to the atomic system.[40,41] Interaction cross-sections are calculated using the dynamic structure factor and energy-loss-function formalism within linear-response theory. The energy-loss function is derived from experimental optical data. Further details of the model are provided in Refs.[40-42]. For $TiO_2$, the energy-loss function, scattering cross-sections, and calculated electronic stopping powers are provided in the supporting information. Comparison with SRIM [43] and CasP 6.0 [44] shows reasonable agreement over the energy range considered. The MC simulations yield distributions of electron and hole energies and densities, as well as the energy transferred to the target atoms.[45] Energy transfer to the lattice is modeled through elastic scattering of electrons and valence-band holes, and through nonthermal atomic acceleration caused by changes in the interatomic potential following electronic excitation.[42,45] The resulting atomic velocity distributions are used as initial conditions for molecular dynamics simulations of the lattice response, performed with LAMMPS.[46] The $TiO_2$ interatomic potential, developed to describe properties of titania polymorphs, is taken from Ref.[47]. It combines Buckingham short-range pairwise interactions with Coulomb forces between partially charged ions. The simulation results are visualized using OVITO.[48]

The crystallization rate was determined from molecular dynamics simulations of solid–liquid two-phase systems. A simulation cell containing crystalline $TiO_2$ in contact with the liquid phase (in total 37026 atoms for anatase and 37632 for rutile) was equilibrated at a selected temperature in range 1800-3200K and zero pressure. The motion of the solid–liquid interface was tracked by monitoring the evolution of the crystalline region using local structural analysis. The crystallization rate was calculated from the time dependence of the interface position, with the growth velocity obtained from the slope of the linear region of the interface displacement versus time. Simulations were performed at several temperatures to determine the temperature dependence of the crystallization rate for rutile and anatase.

Solid and liquid free energies were calculated using nonequilibrium thermodynamic integration methods in system containing 16200 atoms for anatase and 16416 atoms for rutile. The solid-phase free energy was obtained using the nonequilibrium Frenkel-Ladd (FL) method, following the approach of Freitas.[38] The system was equilibrated for 20 ps before thermodynamic switching to Einstein crystal, which was performed over 50 ps in each direction. The spring constant was set to $k$=7 eV/Å$^2$, based on tests with different values of $k$, to ensure system stability across the full temperature range studied. FL calculations were carried out from 1800 to 3200 K in 200 K increments and zero pressure. The liquid-phase free energy was calculated using the Uhlenbeck–Ford (UF) model as a reference system, following the approach of Leite.[39] The simulations contained 16416 atoms and included 20 ps of equilibration followed by 100 ps of switching, which was performed linearly in time. The UF potential parameters were $p = 50$, $\sigma = 1.5$ Å, and a cutoff radius of $r_c = 5\sigma$.

## Acknowledgements

This work was financially supported by the National Natural Science Foundation of China (Grant Nos. 12575301, 12405320, 11905262), and the Natural Science Foundation of Gansu Province (Grant No. 25JRRA463), and the CAS President's International Fellowship Initiative (Grant Nos. 2026PVB0105, 2024PVB0001). The work of R.A. Rymzhanov was funded by the Russian Science Foundation, the Russian Federation (grant No 25-72-10101, https://rscf.ru/project/25-72-10101/). We would like to thank the accelerator staff of the HIRFL (31111.02.HIRFL) for providing the high quality of heavy ion beams. This work has been carried out using computing resources of the Federal collective usage center Complex for Simulation and Data Processing for Mega-science Facilities at NRC "Kurchatov Institute", http://ckp.nrcki.ru//. The research is carried out using the equipment of the shared research facilities of HPC computing resources at Lomonosov Moscow State University.

## Conflicts of Interest

The authors declare no conflicts of interest.

## Data Availability Statement

The data that support the findings of this study are available from the corresponding authors upon reasonable request.

**Table of Contents（TOC）**

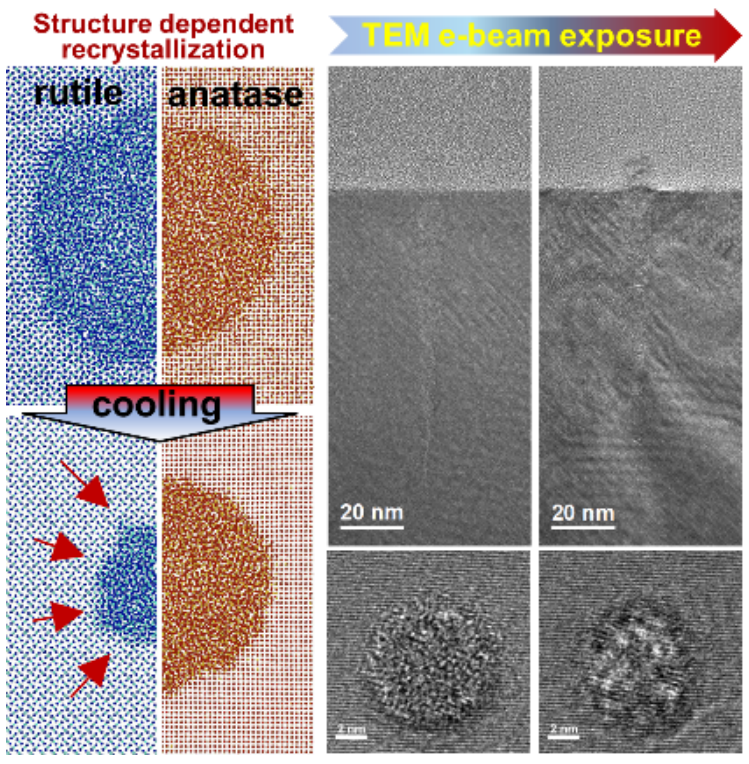

Complementary TEM and molecular dynamics study reveals crystal structure dependent distinct recrystallization pathways in rutile and anatase $TiO_2$, governing swift heavy ion track morphology. Rutile forms hourglass-shaped tracks, while anatase exhibits cylindrical damaged regions. Electron beam exposure promotes track and hillock recrystallization, highlighting the need to account for beam-induced effects in TEM characterization.